\RequirePackage{lineno}
\documentclass[prl, etal, twocolumn,  longbibliography, amsmath,superscriptaddress,amssymb]{revtex4}
\usepackage{graphicx}% Include figure files
\usepackage{subfigure}
\usepackage{emf}
\usepackage{dcolumn}% Align table columns on decimal point
\usepackage{bm}% bold math
\usepackage{verbatim}
\usepackage{xcolor}
\usepackage{braket}
\usepackage{tikz}
\usetikzlibrary{shapes.geometric, arrows}
\usepackage{pgfplots}
\usepackage{mathtools}
\usepackage{color}
\usepackage{balance}
\usepackage{lineno}
\usepackage{upgreek}

\usepackage{CJK}

\usepackage{makeidx}
\begin{document}

%\linenumbers
%\preprint{}
%\begin{CJK*}{UTF8}{gbsn}

\title{Ultrashort Time-Integrated Diagnosis of Laser-Heated Deuterium Ions in Dense Plasma via Fusion Neutron Spectra}

\def\fdu {Key Laboratory of Nuclear Physics and Ion-beam Application (MoE), Institute of Modern Physics, Fudan University, Shanghai 200433,  China}
\def\sjtu {Key Laboratory of Laser Plasma (MoE), School of Physics and Astronomy, Shanghai Jiao Tong University, Shanghai 200240, China}
\def\ifsa {IFSA Collaborative Innovation Center, Shanghai Jiao Tong University, Shanghai 200240,  China}
\def\LOP{Beijing National Laboratory of Condensed Matter Physics, Institute of Physics, Chinese Academy of Sciences, Beijing 100190, China}
\def\SARI{Shanghai Advanced Research Institute, Chinese Academy of Sciences, Shanghai 201210, China}
\def\UCAS{School of Physical Sciences, University of Chinese Academy of Sciences, Beijing 100049, China}
\def\SSLake{Songshan Lake Materials Laboratory, Dongguan 523808, China}

\author{Jie Feng $^a$} \affiliation{\sjtu}\affiliation{\ifsa}\
\author{Hao Xu $^a$}   \affiliation{\sjtu}\affiliation{\ifsa}
\author{Mingxuan Wei}   \affiliation{\sjtu}\affiliation{\ifsa}
\author{Mingyang Zhu}   \affiliation{\sjtu}\affiliation{\ifsa}
\author{Xichen Hu}   \affiliation{\sjtu}\affiliation{\ifsa}
\author{Bingzhan Shi}   \affiliation{\sjtu}\affiliation{\ifsa}
\author{Fuyuan Wu}   \affiliation{\sjtu}\affiliation{\ifsa}
\author{Weijun Zhou}   \affiliation{\sjtu}\affiliation{\ifsa}
\author{Wenchao Yan}   \affiliation{\sjtu}\affiliation{\ifsa}
\author{Guoqiang Zhang}  \affiliation{\SARI}
\author{Jinguang Wang}  \affiliation{\LOP}
\author{Yifei Li}  \affiliation{\LOP}
\author{Xin Lu}  \affiliation{\LOP}\affiliation{\UCAS}\affiliation{\SSLake}
\author{Liming Chen} \email[Corresponding author:] {lmchen@sjtu.edu.cn}\affiliation{\sjtu}\affiliation{\ifsa}
%\author{Jie Zhang} \affiliation{\sjtu}\affiliation{\ifsa}

\date{\today}

\begin{abstract}
	
The ultrashort time-integrated diagnosis of ions plays a vital role in high energy density physics research. However, it is extremely challenging to measure in experiment. Here, we demonstrate a reliable approach for investigating the dynamics of deuterium ions in dense plasma. By irradiating a heavy water stream with the hundred Hertz repetitive intense femtosecond laser pulses, the neutrons from $D(D, n)^3He$ reaction can be detected via a single Time-of-Flight detector to accumulate the spectrum with a fine energy-resolution. This spectrum has been utilized to calculate the temperature and angular distribution of deuterium ions transported in plasma. And the calculated results are well verified by particle-in-cell simulations of deuterium ions dynamics.	Our method paves a new way for diagnosing ions picoseconds time-integrated dynamics in plasma and holds great potential for understanding the ions transport process in high-energy density matters and studying laser plasma ion acceleration.

\end{abstract}
\maketitle

\paragraph{Introduction.}

%% 1st paragraph
High energy density physics (HEDP) is an emerging frontier interdisciplinary field and a significant branch of physics. It primarily focuses on investigating the properties and motion laws of matter under conditions where the energy density exceeds 0.1 million J/cm$^3$ \cite{HEDP-bk,HEDP-nf}. The HED states can be driven by intense lasers \cite{HPL,Ren-PRL}, Z-pinch \cite{HEDP-laser,Z-pinch}, and other methods \cite{HEDP-ion,HEDP-sun}. Generally, high-power lasers can achieve states with energies ranging from 100s to 1000s eV and densities of 10$^{21}$ to 10$^{24}$ cm$^{-3}$ \cite{HEDP-region}. The ultrashort time-integrated dynamics of ions in such a HED environment is of crucial importance for HEDP studies. However, it is extremely challenging to measure experimentally. Although there are works that utilize high-resolution X-ray spectroscopy to diagnose the temperature and density of ions \cite{Xray-PR,Xray-td-HPL,Xray-tem-den}, it is difficult to provide other specific information such as angular and energy distributions. These in-depth dynamic details can be readily obtained through particle-in-cell (PIC) simulations \cite{PIC}, but they often deviate significantly from general experimental measurements \cite{nToF-TNSA}, mainly due to the integration over a long period of time and over a large-scale area. Therefore, short time-integrated and localized measurement is particularly crucial for providing reliable ion dynamics in the target.

%% 2nd paragraph
The HED environment naturally furnishes conditions for triggering nuclear fusion reactions that generate fast neutrons \cite{nano-fusion,collide-fusion,cluster-fusion,Dit-fusion}. The angular distribution and energy spectrum of these neutrons are closely tied to the dynamics of laser-heated ions \cite{Ed-En}. In recent years, neutron spectra have been successfully utilized to study laser plasma ion acceleration and estimate the time duration of femtosecond laser-driven nuclear fusion \cite{nToF-TNSA,BOA,fusion dynamics}. However, due to the constraint of energy-resolution \cite{kHz-D2O Hah,GQzhang-PLA,WWZ-PLB,kHz-D2O knight}, it is nearly impossible to obtain dynamic information of ions from neutron spectra.

%% 3rd paragraph
Conventionally, fast neutron spectra are obtained through proton-recoil and nuclear activation methods \cite{proton recoil, nuc act}, which typically exhibit poor energy-resolution ($\sim$500 keV at 3 MeV) and involve complex data processing. Another widely employed approach is the Time-of-Flight (ToF) measurement \cite{nToF}, which offers better convenience and precision. In practice, however, neutrons hitting the ToF scintillator detector generate a pulse signal whose amplitude depends not only on their count or energies \cite{RSI calibration}, making it difficult to acquire an accurate neutron spectrum. Additionally, a single neutron signal pulse has a long decaying tail ($>$10s ns), leading to the loss of many details in the energy spectrum directly converted from the overlapped neutrons ToF spectrum \cite{kHz-D2O Hah,ToF-HPL}. Thus, in inertial confinement fusion experiments, a large array ToF detector in the single-neutron-counting (SNC) mode is typically used to obtain the precise energy spectrum, which can reveal the ion temperature in the fusion capsule \cite{nToF}. Nevertheless, due to the detector covering a large solid angle, this method is only applicable in cases of symmetrical angular distribution.

%% 4th paragraph
In this letter, a single ToF detector in SNC mode is employed to measure the neutron spectrum with an energy-resolution of 100 keV at 3 MeV through high repetition-rate femtosecond laser irradiation of a heavy water stream. Utilizing the precise experimental spectrum, the dynamics of laser-heated deuterium ions in heavy water plasma predicted by PIC simulation are successfully confirmed. Subsequently, a simple numerical model is utilized to directly solve the ions dynamics, which also shows excellent agreement with the PIC simulation results. Through this novel approach, neutrons from nuclear fusion can serve as a powerful tool for diagnosing ion dynamics in the HED environment.

\paragraph{Experimental setup and results.}

\begin{figure} [b]
	\centering
	\includegraphics[width=0.49\textwidth]{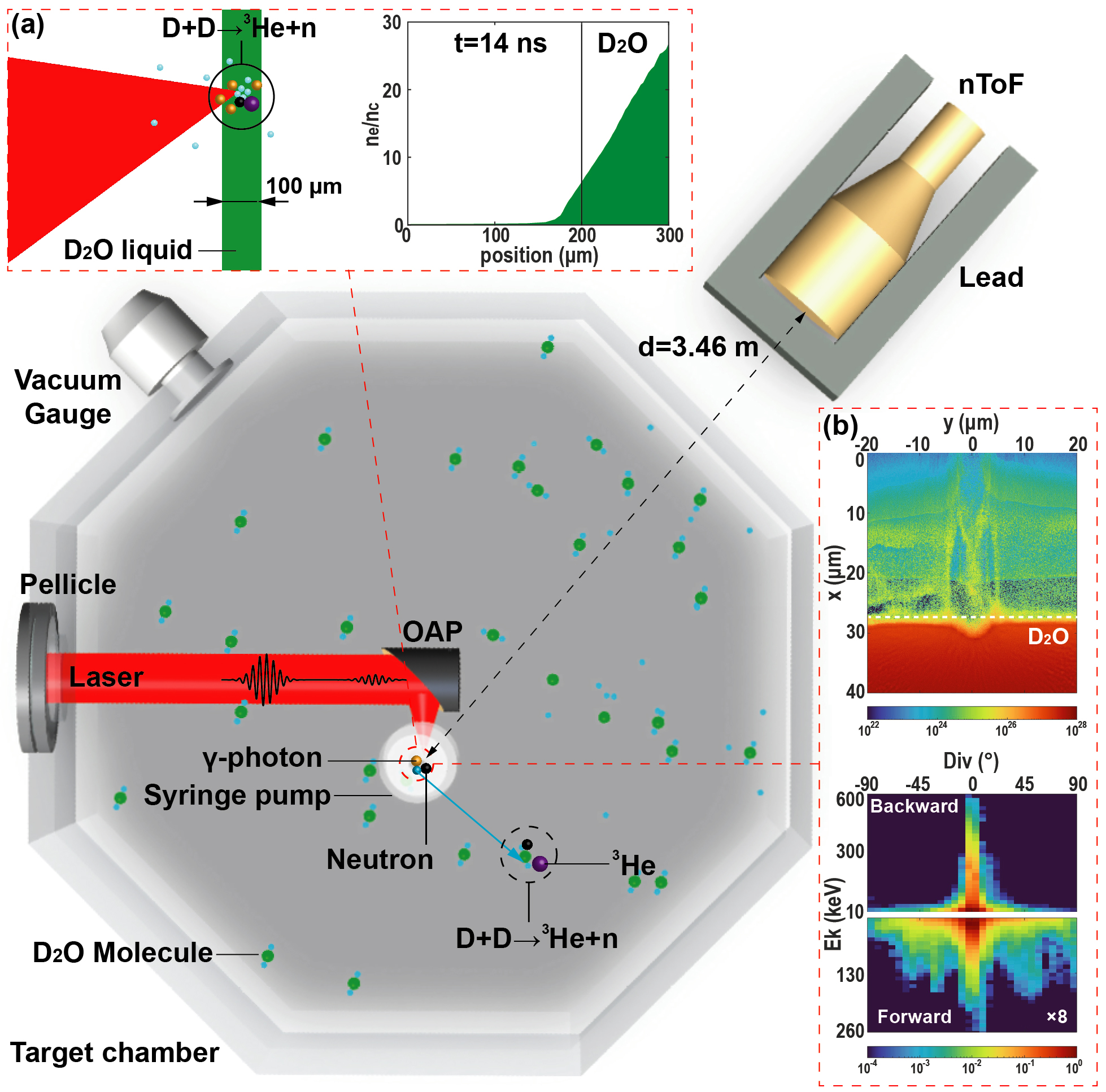}
	\caption{Experiment setup. (a) Fluid simulated pre-plasma density distribution. (b) PIC simulated deuterium ions density distribution and energy /angular distributions at 4 ps.}
	\label{fig:exp_setup}
\end{figure}

%% 1st paragraph
The experiment was conducted using a Ti: sapphire laser system at the SECUF within the Institute of Physics, Chinese Academy of Sciences. The experimental setup is depicted in Fig. 1. A linearly polarized laser pulse with a repetition rate of 100 Hz, a center wavelength of $\sim$800 nm, a duration of 20 fs, and an energy of 20 mJ was focused onto a $w_0=1.2$ $\mu$m spot by means of an f-number 1.3 off-axis parabolic mirror, resulting in a peak intensity of $I\sim2\times10^{19}$ in vacuum. An undisturbed 100 $\mu$m diameter D$_2$O stream at room temperature was delivered through an 80 $\mu$m diameter capillary by a syringe pump and allowed to fall into a reservoir. During the experiment, the evaporated water filled the chamber to a pressure of 2000$\sim$3000 Pa. The vapor pressure of water not only provided a deuteron catcher but also prevented the target stream from freezing. A pellicle was employed to prevent damage to optical components in other vacuum chambers. The laser focus was set $\sim$200 µm below the tip of the capillary and at the front surface normal to the flowing stream.

%% 2nd paragraph
In order to enhance the absorption of laser pulse energy, a pre-pulse was deliberately released from the regenerative amplifier to generate a pre-plasma in the vicinity of the stream. The pre-pulse arrived 14 ns before the main pulse with an intensity contrast ratio of $\sim 10^4$. The laser system has an amplified-spontaneous-emission intensity contrast ratio of $\sim 10^8$. When the main pulse arrived, the ionized pre-plasma expanded into a lower density distribution, as depicted by the 1D Multi-IFE simulation in Fig. 1(a) \cite{Multi, HEDP-laser}. The main pulse was focused onto the pre-plasma and transferred energy to electrons, then these electrons heated a part of deuterium ions. These energetic deuterium ions have a certain probability of triggering fusion reactions with deuterium ions in the plasma. Moreover, some of these energetic deuterium ions would escape from the laser focus through Coulomb explosion, and have a certain probability of triggering fusion with deuterons of D$_2$O vapor in the target chamber. The neutrons generated from the fusion reaction were detected by a liquid scintillator EJ-301 detector \cite{RSI calibration,EJ 301}, which was set 135$^\circ$ away from the direction of laser propagation, at a distance of 3.46 m.

\begin{figure} [h]
	\centering
	\includegraphics[width=0.44\textwidth]{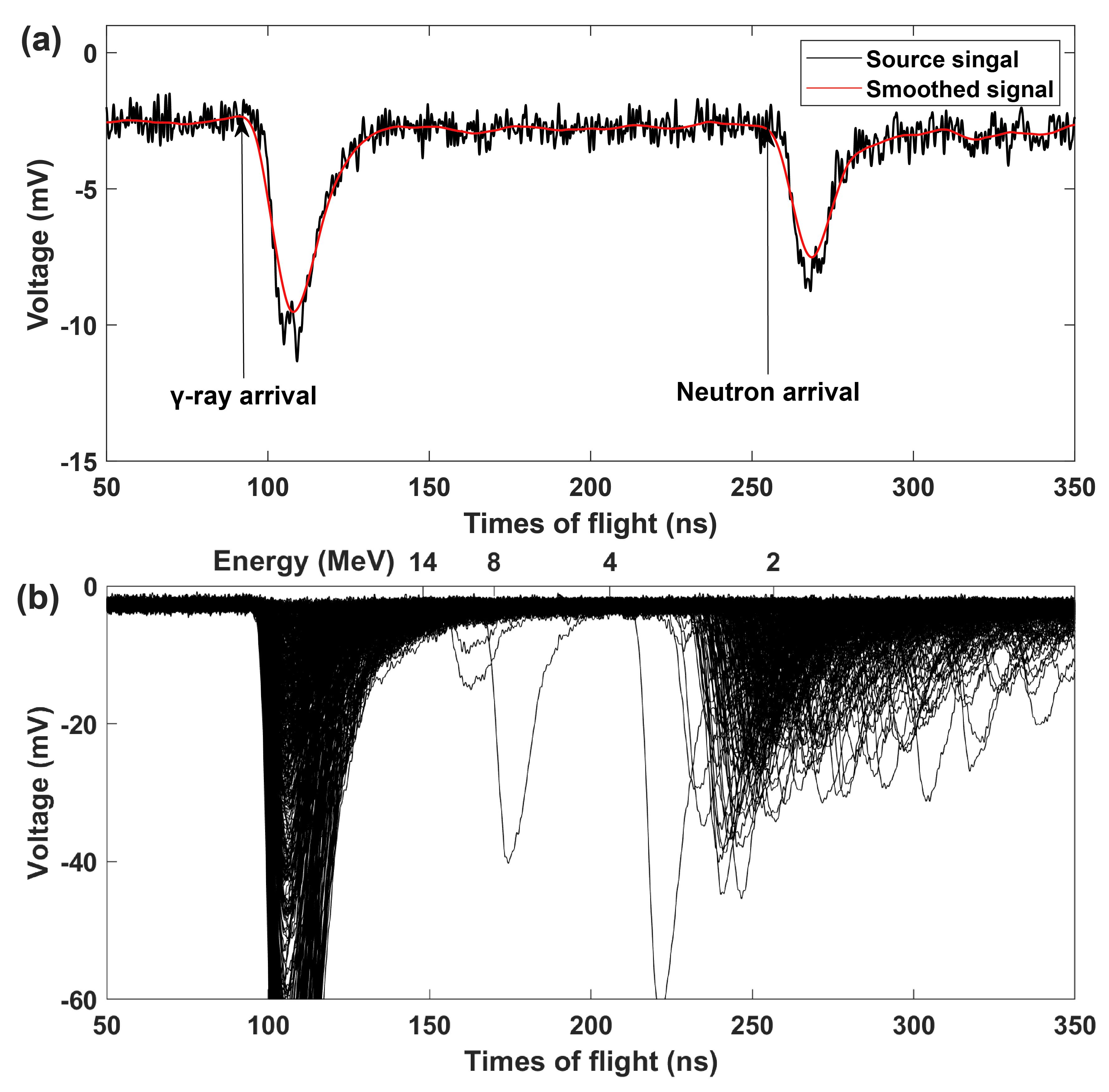}
	\caption{Neutron Time-of-Flight spectra. (a) Typical single-neutron-counting spectrum. (b) A set of spectra for 10 mins in 100 Hz.}
\end{figure}

%% 3rd paragraph
The typical SNC ToF is shown in Fig. 2(a), the starting point of the two pulses are the moment when the $\gamma$-ray/neutron reach the detector respectively. The neutron kinetic energy $E_n$ can be calculated based on the time difference between the two arrival moments $\Delta t$ and the detection distance $L$, as $E_n=\frac{1}{2}\cdot m_n\cdot (\frac{L}{\Delta t+L/c})^2$, where $m_n$ is the neutron mass. During the experiment, the ToF detector keeps recording at most one neutron for each count, as shown by the signals in Fig. 2(b).

\begin{figure} [b]
	\centering
	\includegraphics[width=0.49\textwidth]{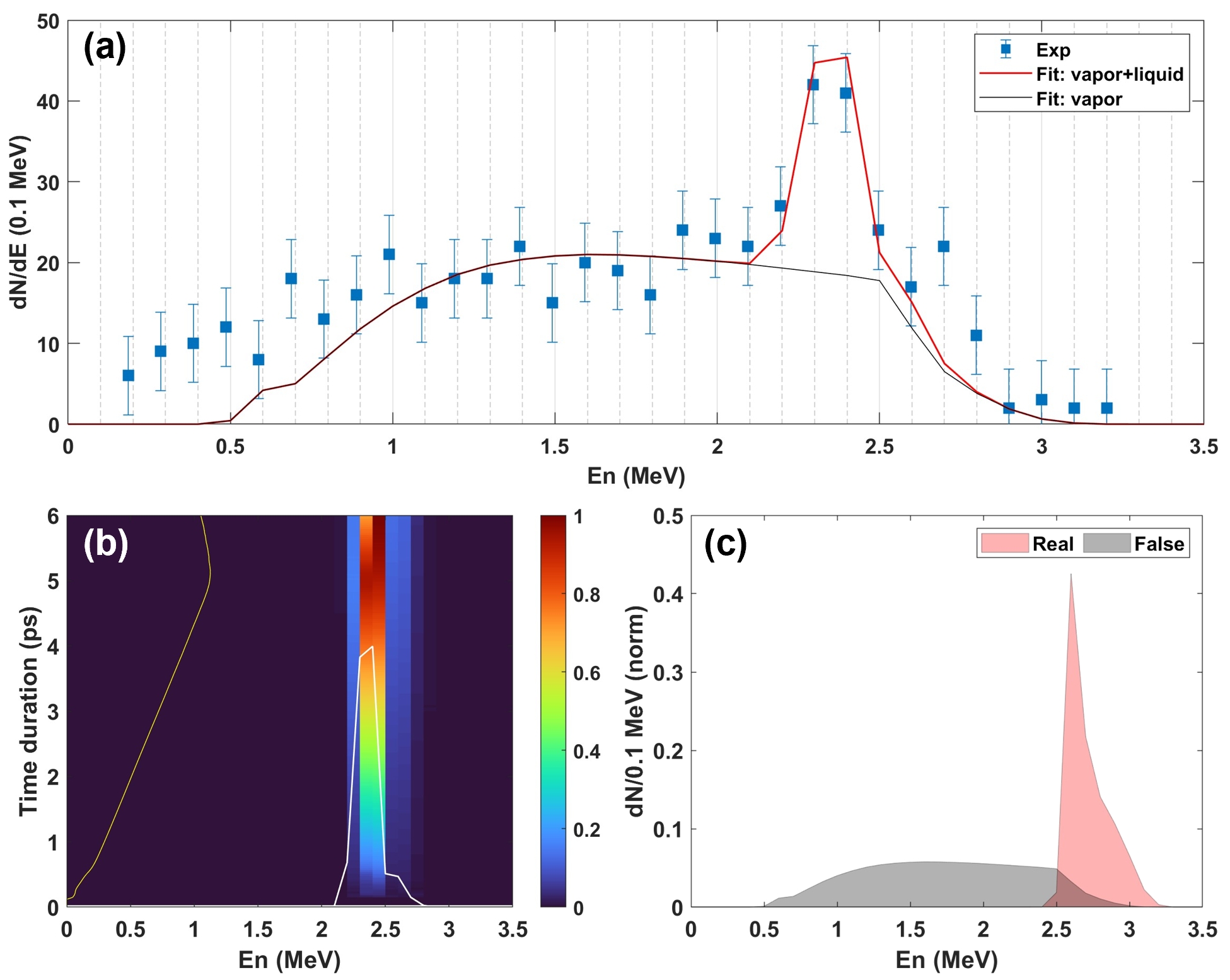}
	\caption{Neutron energy spectra. (a) Neutron spectra from the sum of stream and vapor, the squares represent experiment data, the errorbar is from standard deviation, and the red/black solid lines corresponds to simulation data. (b) Time evolution of the neutron spectra from D$_2$O stream at different moments, and the white line is the sum spectrum. (c) Neutron spectra from D$_2$O vapor. The gray spectrum considers the time delay effect, and the red spectrum is the real energy distribution.}
\end{figure}

%% 4th paragraph
The energy range of the neutrons generated from nuclear reaction $A(a,n)B$ can be estimated by the following reaction energy $Q$-equation \cite{Ed-En},
\begin{align} 
Q=(1+\frac{m_a}{m_B})E_n-(1-\frac{m_a}{m_B})E_a-\frac{2\sqrt{m_am_nE_aE_n}}{{m_B}/cos(\theta)},
\end{align}
where $\theta$ is the angle between the emitted neutron and the incident nucleus $A$. For $D(D,n)^3He$ reaction with the $Q$-value of approximately 3.27 MeV, considering that deuterium ions have a maximum energy of 1 MeV in $4\pi$, the neutron energy range is about 2 to 4 MeV. However, many recorded neutrons are located in the lower energy range. This is caused by the escaped laser-heated deuterium ions triggering nuclear fusion in the environment D$_2$O vapor. These deuterium ions have traveled a long distance at slow velocities ($\sim10^6$m/s) before fusion, resulting in the generated neutrons arriving at the ToF detector later. According to Fig. 2(b), the neutron energy spectrum can be counted and is shown in Fig. 3(a). The energy-resolution is estimated to be 100 keV at 3 MeV, considering the detection distance, scintillator thickness and time deviation for reading the ToF. In addition, there are several higher energy neutrons, which are generated by the $T(D,n)^4He$ reaction with the $Q$-value of approximately 17.59 MeV. The tritium nuclei with an energy of approximately 1 MeV are generated from the $D(D, p)T$ nuclear reaction.

\paragraph{Analysis.}

%% 1st paragraph
According to Equation (1), the energy of the detected neutron is closely associated with the energy and angular distribution of deuterium ions. To gain a profound understanding of the instantaneous dynamics of laser-heated deuterium ions, we conducted a 2D-PIC simulation using the EPOCH code \cite{PIC,Morris2022}. The pre-plasma density distribution input for the PIC simulation, which is caused by the laser pre-pulse, is provided by the fluid simulation via the 1D-Multi code. The size of the PIC simulation box is  $50\times50$ $\rm\mu m^2$, with $1875\times1875$ cells in the x and y directions. The p-polarized laser main pulse propagates along the x-direction, and the laser-focusing plane is located at x = 30 $\mu$m.

%% 2nd paragraph
The laser main pulse transfers energy to electrons through mechanisms such as vacuum heating, resonance absorption or $J\times B$ heating etc \cite{heating regime, ion-vf, ion-vf-sim}. Subsequently, the hot electrons penetrate into the dense plasma and heat deuterium ions. The PIC simulation results of deuterium ions at 4 ps are presented in Figure 1(b). When the laser pulse reaches the critical density surface at approximately x = 27 $\mu$m, it is reflected and drives a large number of electrons into the vacuum, forming a target normal sheath electric field on the order of TV/m near the surface \cite{efield-TNSA,Ion-acc}. The backward accelerated deuterium ions have a higher maximum energy of approximately 600 keV, which is about three times that of the forward deuterium ions. Due to the toroidal quasi-static magnetic field sustaining the associated electrostatic field, which contributes to collimation \cite{Bfield-ion,Collimation-ion}, the backward deuterium ions have a small divergence angle. It is feasible to measure this distribution by general ion detectors, as well as the ions$'$ energy distribution. However, the forward heated deuterium ions in the dense plasma have a more complex dynamic distribution and cannot be diagnosed.

%% 3rd paragraph
To experimentally verify the dynamic information of deuterium ions in plasma, we utilized the ion dynamics simulated by PIC as input to calculate the neutron spectrum of the $D(D,n)^3He$ nuclear reaction. Subsequently, we compared it with the experimental neutron spectrum. As depicted in Figure 1, neutrons are generated from the D$_2$O stream and the vapor within the target chamber. In the case of the \textbf{stream}, the thin target approximation is employed to calculate the neutron energy spectrum. Considering that the change in the cross-section of the $D(D,n)^3He$ reaction is negligible within an extremely short time step. For example, when $E_D=300$ keV and $dt=30$ fs, the transport distance $l_d$ is approximately 150 nm, the energy loss is around 6 keV, and the cross-section change is approximately 1.6\%. Moreover, its projective range is approximately 5 $\mu$m, and the propagation time in the D$_2$O stream is approximately 1.5 ps. By using the energies and momenta of deuterium ions simulated by PIC as input for calculating the neutron spectrum, the neutron yield of a single macro-particle within a time step can be expressed as follows,

\begin{align}
Y_n=w\cdot\sigma_\Omega(\theta_n,E_D)\cdot n_D(x)\cdot v_D\cdot dt,
\end{align}
where $w$ is micro-particle weight, $\sigma_\Omega$ is differential cross section calculated by DROSG code \cite{sig-cs}, $n_D$ is background deuteron density, $v_D$ is incident deuterium ions velocity, and $\theta_n$ is the included angle between incident deuterium ion and neutron detection direction. The corresponding neutron energy can be expressed as follow according to Eq (1),
\begin{equation}
\begin{split}
\sqrt{E_n}=\ & 0.3535\cdot cos(\theta_n)\cdot \sqrt{E_D}\\&
+\sqrt{(0.125\cdot cos^2(\theta_n)+0.25)\cdot E_D+2.475}.
\end{split}
\end{equation}
Subsequently, the neutron spectrum can be counted by analyzing the neutron yield and corresponding energy contributed by the deuterium ions with kinetic energy $E_D>10$ keV, momentum $p_x>0$ (forward), $i.e.$ $(E_n,Y_n)=hist(E_{ni},Y_{ni})$, $i$ is the tag of deuterium ions. The neutron spectrum evolution is shown in Fig. 3(b), and the white line represents the spectrum of total generated neutrons.

%% 4th paragraph
In the case of \textbf{vapor}, the thick target approximation is employed, considering the ions have a long-projected range ($e.g. \sim$20 cm at 300 keV). Assuming the projected range $l_j$ for one micro-particle with original energy $E_0$ [keV] losing every 1 keV energy, it can be expressed as $l_j=\frac{dx}{dE}(E_0-(j-1))$, where $j$ is the step number, $\frac{dE}{dx}(E)$ is the stopping power calculated by the SRIM-2013 code \cite{srim}. The neutron yield of a single micro-particle within a step can be expressed as
\begin{equation}
	Y_{nj}=w\cdot \sigma_\Omega(\theta_j,E_{j-1})\cdot n_{vapor}\cdot \frac{dx}{dE}(E_{j-1}),
\end{equation}
where $\theta_j$ is the revised included angle by the projected distance $\sum_{j=1}^{j}l_j$ to the starting position, $n_{vapor}$ is the deuteron density in vapor. The corresponding neutron energy can be calculated by Eq (3). The energy spectrum is acquired from $(E_n,Y_n)=hist(E_{nji},Y_{nji})$ for $E_D>10$ keV, $p_x<0$ (backward), and the result is shown by the red shadow in Fig. 3(c).

%% 5th paragraph
However, the neutrons generated in vapor environment experience a time delay when reaching the ToF detector, due to the propagation distance of deuterium ion from TNSA. Thus, the neutron energy calculated from the ToF detector can be expressed as follows,
\begin{equation}
	E_{nj}=\frac{1}{2}m_n\left(\frac{l}{\sum_{j=1}^{j}\frac{l_j}{\sqrt{2E_{j-1}/m_D}}+\frac{d_j}{\sqrt{2E_{j-1}/m_n}}}\right)^2,
\end{equation}
where $m_D$ is deuterium ion mass, $d_j$ is the distance from the neutron generated position to the ToF detector. The revised spectrum on the detector is shown by the gray shadow in Fig. 3(c). The real energy range of neutrons from vapor is 2.5 to 3.3 MeV. However, the range detected by the detector is 0.5 to 3 MeV.

%% 6th paragraph
The experimental energy spectrum is well-fitted by the sum of calculation spectra derived from PIC simulation. As depicted in Fig. 3(a), the spectrum can be divided into two parts. Notably, the neutron yield contributed by backward energetic deuterium ions is approximately six times higher than that of the forward ones. It is worthy of mention that the energy range of 2 to 3 MeV can unveil the dynamics of deuterium ions moving forward in dense plasma. Additionally, this energy range has been excellently fitted by experimental results, indicating that the energy and angular distribution of the PIC-simulated deuterium ions are experimentally demonstrated.

\paragraph{Discussion.}

%% 1st paragraph
To directly obtain the dynamics of energetic deuterium ions in dense plasma based on an experimentally acquired high-resolution neutron spectrum, a simple model is employed. Assuming that these ions possess an energy spectrum following the Maxwell-Boltzmann distribution $f_{MB}(E_D,T_D)$ and an angular distribution following the Lorentz distribution $f_L(\theta,\gamma)$, these ions trigger nuclear fusion in D$_2$O stream plasma to generate neutrons. The neutrons have energies as determined by Eq (3), and the yield is calculated via the following formula,
\begin{equation}
Y_n(E_D,\theta_n)=\int_{E_1}^{E_2}\int_{\theta_1}^{\theta_2}f_{MB}\cdot f_L\cdot\frac{n_D\cdot \sigma_\Omega}{\frac{dE_D}{dx}(E_D)}\cdot dE_D\cdot d\theta_n;
\end{equation}
Consequently, the temperature and angular distributions of these ions can be determined by fitting the neutron energy spectrum to the experimental spectrum. The method is described as follows:

By fitting the temperature $T_D$ and distribution width factor $\gamma$, we ensure that the fitted spectra are as close as possible to the experimental spectrum via evaluating Chi-Square value, as shown in Fig. 4(a). Limited by the energy-resolution 100 keV of the spectrum, this approach can only determine a temperature range of 20 to 30 keV. As a demonstration, the PIC-simulated temperature of deuterium ions in dense plasma is precisely within this temperature range, as depicted in Fig. 4(b). The angular distribution of the ions is shown by the gray band in Fig. 4(c), which is close to the PIC-simulated distribution (shown as the red line). Of course, there is still a deviation within the range of 2.6 to 2.8 MeV, which could be contributed by either forward or backward deuterium ions. If this portion of neutrons comes from the forward ions, it indicates that in the experimental situation, in addition to the energy and angular distributions shown in the PIC simulation, there should also be a portion of deuterium ions with energy greater than 30 keV and an emission angle between 50$^\circ$ and 90$^\circ$, according to the neutron energy colormap as shown in Fig. 4(d). It is precisely because the nuclear fusion occurring in the dense plasma has an extremely short duration and the generated neutrons can carry out the dynamic information of deuterium ions without distortions. Enabling the use of high-resolution neutron energy spectrum to demonstrate ion ultrashort time-integrated temperature and angular distributions in plasma.   
\begin{figure} [h]
	\centering
	\includegraphics[width=0.49\textwidth]{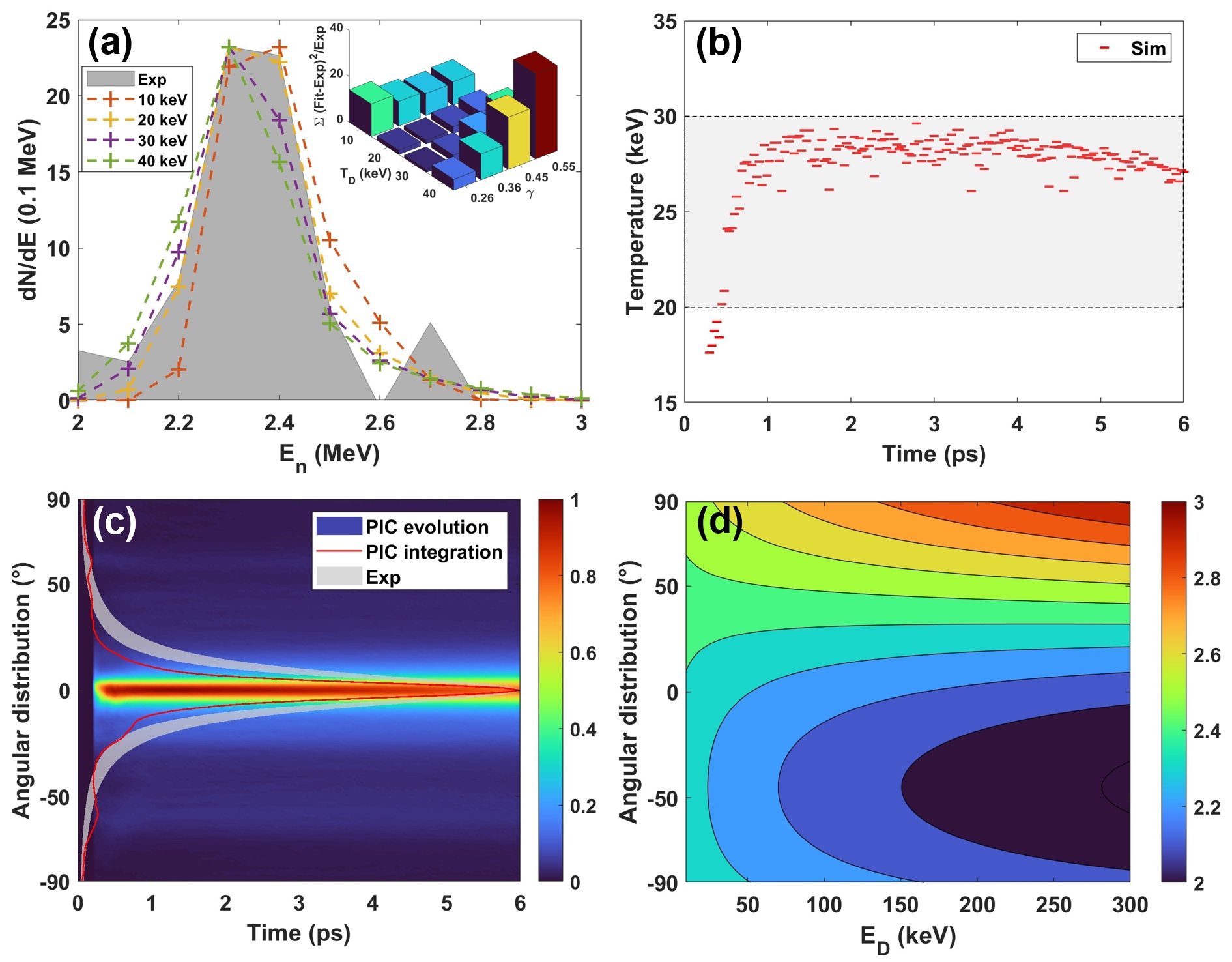}
	\caption{Deuterium ions dynamics in dense plasma. (a) Neutron energy spectra fitting for the cases of different deuterium ion temperature. The gray region is the experimental neutron spectrum from the D$_2$O stream plasma. The illustration is Chi-Square value. (b) Deuterium ions temperature in dense plasma, and the short red lines are PIC simulated temperatures at different times. (c) Deuterium ions angular distributions, the red line is the simulated average distribution, the gray band corresponds to the range of 20 to 30 keV. (d) Relationship between neutron energy and ion energy-angular distribution.}
\end{figure}

\paragraph{Conclusion.}
We have demonstrated a novel approach of studying the dynamics of deuterium ions in dense plasma based on high repetition rate fs laser irradiation D$_2$O stream. Firstly, the experimental spectrum obtained by a single ToF detector in SNC mode with fine energy-resolution 100 keV at 3 MeV is in excellent agreement with the simulated spectrum. This indicates that the PIC simulation result can accurately reflect the ultrashort time-integrated dynamics of deuterium ions in the experiment. Subsequently, the experimental neutron spectrum is also well fitted by the calculated spectrum using a short time-integrated model based on deuterium ions with a temperature of 20 to 30 keV and an emission angle following a Lorentz distribution. These results are consistent with the PIC simulation outcome. Therefore, our work paves a new way for diagnosing the dynamics of ions in dense plasma via only a high-resolution neutron spectrum. This is of great significance for studying the transport process of ions in HED environments and the regime of laser plasma ion acceleration.

\begin{acknowledgments}
This work was supported by the National Natural Science Foundation of China (12335016, W2412039, 11991073, 12305272, 11721404, 12074251, 12235003, U2330401, 12088101), the Strategic Priority Research Program of the CAS (XDA25030400, XDA25010100, XDA25010500, XDA25051200), the National Key R\&D Program of China (2021YFA1601700). This work was carried out at the Synergetic Extreme Condition User Facility (SECUF, http://cstr.cn/31123.02.SECUF.D3).

$^a$ J.F. contributed equally to this work with H.X..
\end{acknowledgments}

%%%% CLEAR DOUBLE PAGE!
%\newpage{\pagestyle{empty}\cleardoublepage}

\end{document}